\documentclass[preprint,showpacs,preprintnumbers,amsmath,amssymb]{revtex4}

\usepackage{graphicx}
\usepackage{dcolumn}
\usepackage{bm}

\def\pmb#1{\setbox0=\hbox{#1}%
  \kern-.025em\copy0\kern-\wd0 
  \kern.05em\copy0\kern-\wd0
  \kern-.025em\raise.0433em\box0 }

\newcommand{\Figuretable}[1]{%
  \begin{center} --------- {\bf #1} --------- \\ \end{center}} 

\begin{document}


\title{
Optimal Fermi-averaging t-matrix for 
($\pi^+$,$K^+$) reactions in $\Lambda$ quasi-free region
}

\author{Toru Harada}%
 \affiliation{%
Research Center for Physics and Mathematics,
Osaka Electro-Communication University, Neyagawa, Osaka, 572-8530, Japan
}

\author{Yoshiharu Hirabayashi}%
\affiliation{%
Information Initiative Center, 
Hokkaido University, Sapporo, 060-0811, Japan
}

\date{\today}

\begin{abstract}

We propose an optimal Fermi-averaging for an elementary 
$\pi^+$+$n$$\to$$K^+$+$\Lambda$ t-matrix under the on-energy-shell 
condition, 
in order to describe ($\pi^+$,$K^+$) reactions on a nuclear target
in the framework of a distorted-wave impulse approximation.
We apply it to calculate $\Lambda$ quasi-free spectra 
from $^{12}$C($\pi^+$,$K^+$) reactions at $p_\pi$= 1.20 GeV/c and 
1.05 GeV/c, and compare them with experimental data.
The results show that the calculated spectra are in 
excellent agreement with the data, 
because the energy-dependence originates from the nature of 
the optimal Fermi-averaging t-matrix.

\end{abstract}

\pacs{21.80.+a, 24.50.+g, 25.80.Pw, 27.20.+n}
\maketitle

\begin{figure}[htb]
  \begin{center}
  \includegraphics[width=0.8\textwidth]{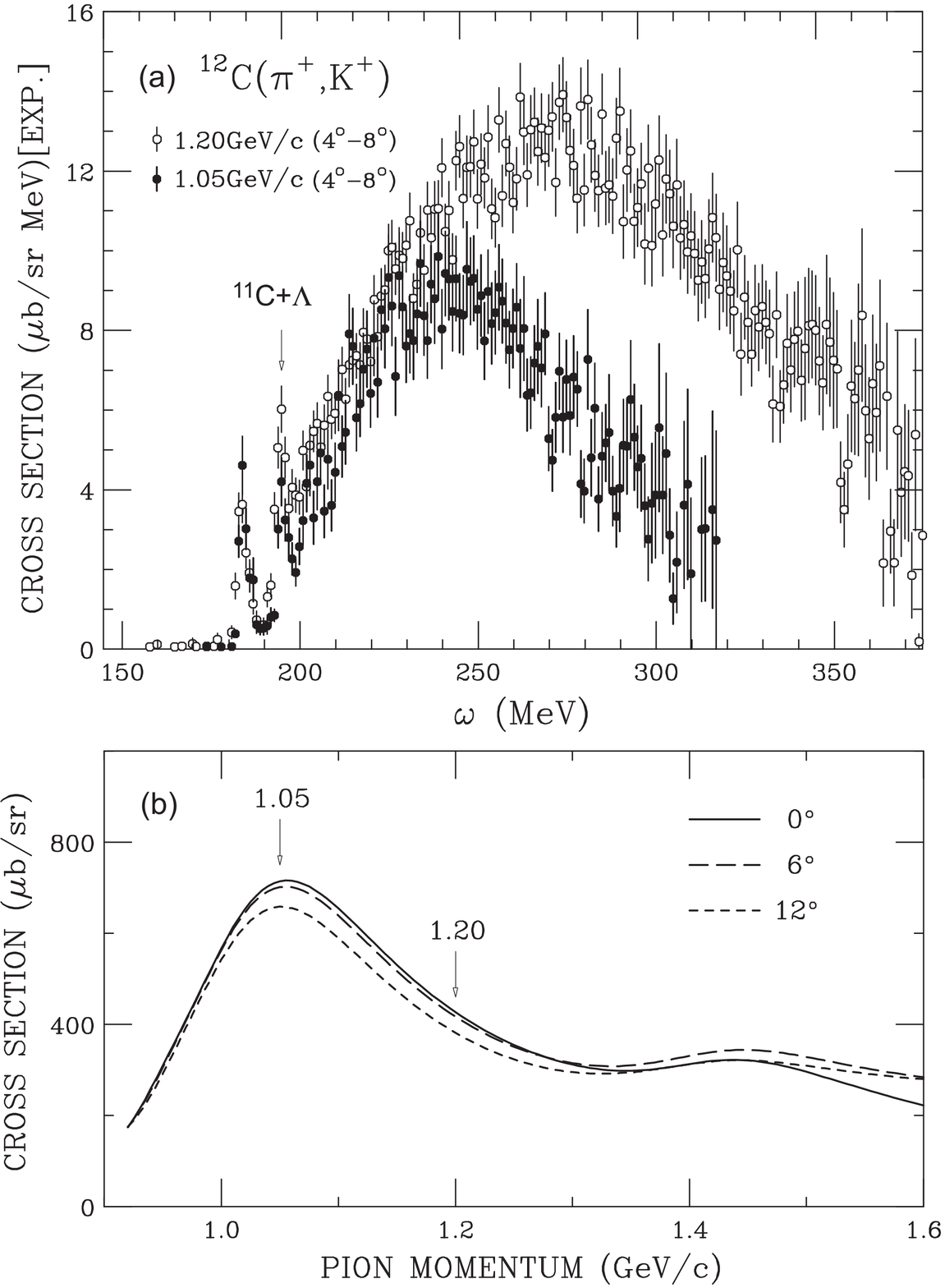}
  \caption{\label{fig:1}
  (a) The ($\pi^+$,$K^+$) data on the $^{12}$C target plot from E438 and E521 
  experiments at KEK \cite{Saha03}, as a function of the energy transfer 
  $\omega$. 
  The open and filled circles denote the cross sections 
  ${\sigma}^{\rm \, exp}_{4^\circ-8^\circ}$ at $p_\pi=$1.20 GeV/c and 
  1.05 GeV/c, respectively. 
  The energy for the $\Lambda$ emitted threshold of 
  $^{11}$C$_{\rm g.s.}$+$\Lambda$ is  194.8 MeV.
  (b) The lab differential cross sections for an elementary $\pi^+ +n \to K^+ +\Lambda$ 
  reaction at $K^+$ forward-direction angles $\theta_{K}=$ 0$^\circ$, 
  6$^\circ$ and 12$^\circ$ \cite{Sotona89}, 
  as a function of the incident $\pi^+$ momentum.   
  }
  \end{center}
\end{figure}

One of the significant subjects in the hypernuclear spectroscopy is 
to elucidate hadronic many-body dynamics with strangeness 
degree of freedom in nuclear and hadron physics. 
In particular, the spectroscopy by ($\pi^+$,$K^+$) reactions
is of great advantage to investigate not only $\Lambda$ deeply-bound 
states but also $\Lambda$ quasi-free (QF) scattering states,
because a momentum transfer to a $\Lambda$ becomes $q \simeq 300 \sim 500$ MeV/c
which goes over a Fermi-momentum $p_F \simeq$ 270 MeV/c 
\cite{Dover80,Motoba88,Hausmann89}.
Recently, observations of $\Lambda$ production 
by $^{12}$C($\pi^+$,$K^+$) reactions were performed 
in E438 and E521 experiments with high quality $\pi^+$ beams at KEK \cite{Saha03}.
In Fig.~\ref{fig:1}a we show the data of the missing mass spectra 
at the incident $\pi^+$ momenta $p_\pi$=1.20 GeV/c and 1.05 GeV/c, 
where the cross sections ${\sigma}^{\rm \, exp}_{4^\circ-8^\circ}$ were obtained 
at a $K^+$ forward-direction angle of $\theta_K=$ 6$^\circ \pm 2^\circ$.
These spectra differ significantly at higher $\Lambda$ excitation energies 
\cite{Saha03}:
(1) a peak position of the QF spectrum at 1.20 GeV/c is 
$\omega \simeq$ 275 MeV which corresponds to about 80 MeV above the 
$^{11}$C$_{\rm g.s.}$+$\Lambda$ threshold, 
while the peak position at 1.05 GeV/c is $\omega \simeq$ 240 MeV
which corresponds to about 45 MeV above the threshold;
(2) width of the QF spectrum at 1.05 GeV/c is about 80 MeV, 
which is narrower than that at 1.20 GeV/c;
(3) a magnitude of the QF peak at 1.05 GeV/c is about 2/3 times 
smaller than that at 1.20 GeV/c.
This distinction would be a slight puzzle, because the lab 
differential cross section for an elementary $\pi^+$+$n$$\to$$K^+$+$\Lambda$ 
reaction, $(d \sigma/d \Omega)^{\rm elem}$, at 1.05 GeV/c 
is quite larger than that at 1.20 GeV/c, 
as shown in Fig.~\ref{fig:1}b \cite{Sotona89}.

\Figuretable{FIG. 1}

In a Fermi gas model \cite{Dalitz76}, 
a strength function for $\Lambda$ production 
is well-known to be characterized 
by the momentum transfer $q$. 
One may attempt to explain the QF spectrum by a change of a momentum 
transfer $q$ as a function of $p_\pi$.
For $p_\pi=$ 1.20 GeV/c, the QF peak which follows $q\simeq$ 400 MeV/c, 
has a position of $\omega^{\rm FG}_{\rm peak} \simeq$  275 MeV 
and width of about 190 MeV,
of which values are similar to these of the data in Fig.~\ref{fig:1}a. 
However, this agreement would be accidental.
Because the ($\pi^+$,$K^+$) reaction is an endothermic reactions, 
the momentum transfer $q$ increases slowly 
as $p_\pi$ decreases toward a $\Lambda$ production threshold \cite{Dover80}.
Then, if we assume $(d \sigma/d \Omega)^{\rm elem}$ 
to be constant, the position of the QF peak should be shifted up and 
the width should be broader as $p_\pi$ decreases. 
For $p_\pi=$ 1.05 GeV/c, the QF peak must have $q\simeq$ 450 MeV/c,
leading to $\omega^{\rm FG}_{\rm peak} \simeq $ 295 MeV 
and width of about 220 MeV. 
It is an opposite tendency of the data. 
This inconsistency brings again the puzzle to our 
attention.
As shown in Fig.~\ref{fig:1}b, there appears a strong 
energy-dependence in $(d\sigma/d\Omega)^{\rm elem}$
due to $N^*$ resonances, 
e.g., S$_{11}$(1680), P$_{11}$(1730) and P$_{13}$(1700) \cite{Sotona89}.
Such an energy-dependence is necessary to describe 
the spectrum in the wide $\omega$ energy-range 
including from $\Lambda$ bound to QF regions, 
and would be one of key elements for solving the puzzle.
Since widths of those resonances are comparable to the range of 
a Fermi-momentum $p_F \simeq$ 270 MeV/c for a nucleon in a nucleus, 
the $\pi^+$+$n$$\to$$K^+$+$\Lambda$ t-matrix must be also Fermi-averaged 
in the nucleus \cite{Allardyce73,Rosenthal80,Dover82}.

In this work, 
we propose an optimal Fermi-averaging for an elementary 
$\pi^+$+$n$$\to$$K^+$+$\Lambda$ t-matrix in 
($\pi^+$,$K^+$) reactions on a nuclear target.
Applying it to a distorted-wave impulse approximation (DWIA), 
we calculate $\Lambda$-hypernuclear QF spectra from
$^{12}$C($\pi^+$,$K^+$) reactions 
at $p_\pi$= 1.20 GeV/c and 1.05 GeV/c,
and compare these spectra with the experimental data. 
We will show that the optimal Fermi-averaging 
t-matrix is essential to describe the $\omega$ energy-dependence 
of the $\Lambda$ QF spectra by the ($\pi^+$,$K^+$) reactions.

\Figuretable{FIG. 2}

\begin{figure}[htb]
  \begin{center}
  \vspace{5.0mm}
  \includegraphics[width=.55\textwidth]{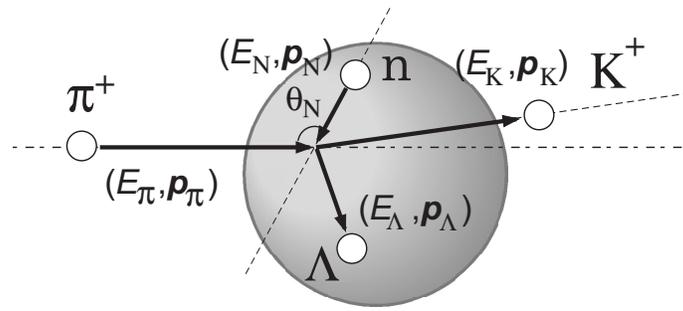}
  \caption{\label{fig:2}
  Impulse approximation for the 
  $\pi^+$+$n$$\to$$K^+$+$\Lambda$ reaction in a nucleus. 
  }
  \end{center}
\end{figure}

Hypernuclear production cross sections have been usually analyzed 
with a DWIA 
\cite{Hufner74,Bouyssy77,Auerbach83,Motoba88,Hausmann89,Morimatsu94}.
In Fig.~\ref{fig:2} we illustrate the $\pi^+$+$n$$\to$$K^+$+$\Lambda$ 
process in a nucleus within an impulse approximation:
After an incident $\pi^+$ with the momentum-energy 
(${\bm p}_\pi$, $E_\pi$) 
interacts with a neutron which has a Fermi-motion with 
(${\bm p}_N$, $E_N$) 
in the nucleus, an outgoing $K^+$ with (${\bm p}_K$, $E_K$) 
and a $\Lambda$ with (${\bm p}_\Lambda$, $E_\Lambda$) are produced.
The total energy and the momentum-energy transfer are 
given as $E_2$=$E_\pi$$+$$E_N$ and 
(${\bm q},\omega$)=(${\bm p}_\pi$$-$${\bm p}_K$,$E_\pi$$-$$E_K$),
respectively.
In the impulse approximation, 
one usually needs an off-energy-shell t-matrix of 
the $\pi^+$+$n$$\to$$K^+$+$\Lambda$ reaction. 
If one inputs experimental data of the elementary process, 
one sometimes replaces it by an on-energy-shell t-matrix
(on-energy-shell approximation).
In this case, when one chooses (${\bm p}_\pi$, $E_\pi$) 
for the incident $\pi^+$ beams as the on-energy-shell, 
the $\pi^+$+$n$$\to$$K^+$+$\Lambda$ t-matrix must become 
constant over (${\bm p}_K$, $E_K$)
which is determined by the 
$\pi^+$+${^{\rm A}}{\rm Z}$$\to$$K^+$+${^{\rm A}_\Lambda}{\rm Z}^*$
kinematics 
not the $\pi^+$+$n$$\to$$K^+$+$\Lambda$ one.

In order to avoid such a procedure and 
to describe the energy-dependence of the ($\pi^+$,$K^+$) 
reaction appropriately, 
we propose an ``optimal'' t-matrix for the 
$\pi^+$+$n$$\to$$K^+$+$\Lambda$ reaction in the nucleus:
We assume that the only process which satisfies 
the on-energy-shell condition in the nucleus, 
i.e., $E_2=E_\pi + E_N = E_K + E_\Lambda$ and 
${\bm p}_\pi+{\bm p}_N={\bm p}_K+{\bm p}_\Lambda$, 
is allowed to contribute to the ($\pi^+$,$K^+$) reaction, 
and do averaging the elementary $\pi^+$+$n$$\to$$K^+$+$\Lambda$
t-matrix over the momentum distribution $\rho{(p_N)}$$\equiv$$|\Psi(p_N)|^2$. 
Once we take (${\bm q},\omega$) which is chosen 
by ${\bm p}_\pi$ and ${\bm p}_K$ on the nuclear target, 
this t-matrix can be calculated as
\begin{widetext}
\begin{equation}
{t}^{\rm opt}(\omega,\theta_K) =
{\displaystyle
  \int^{\pi}_{0} \sin{\theta_N} d \theta_N \int_{0}^{\infty} 
  dp_{N} p_N^2 \rho{(p_N)}
  {t}(E_{2};{\bm p}_\pi,{\bm p}_N) 
\over 
  \displaystyle
  \int^{\pi}_{0} \sin{\theta_N} d \theta_N \int_{0}^{\infty} 
  dp_{N} p_N^2 \rho{(p_N)}
  }\Biggl|_{{\bm p}_N={\bm p}^*_N},
\label{eqn:e5}
\end{equation}
\end{widetext}
where $\cos{\theta_N}$=${\hat{\bm p}}_\pi\cdot{\hat{\bm p}}_N$. 
The momentum ${\bm p}^*_N$ is a solution for the ``on-energy-shell'' equation 
\begin{equation}
\omega=\sqrt{({\bm p}^*_N+{\bm q})^2+m_\Lambda^2}
-\sqrt{{{\bm p}^*_N}^{2}+m_N^2},
\label{eqn:e4a} 
\end{equation}
which connects to the nuclear kinematics $\omega = E_f- E_i$, 
where $E_f$ and $E_i$ ($m_\Lambda$ and $m_N$)
are energies of a hypernuclear final state and 
a target-nuclear initial state
(masses of a $\Lambda$ and a neutron), respectively. 
We regard the r.h.s. in eq.(\ref{eqn:e5}) 
an ``optimal Fermi-averaging'' under the on-energy-shell condition.
As a result, the optimal Fermi-averaging t-matrix 
${t}^{\rm opt}(\omega,\theta_K)$ acquires a dependence on $\omega$.

The double-differential production cross section for the 
($\pi^+$,$K^+$) reaction at a $K^+$ forward-direction angle 
$\theta_{K}$ in the lab frame is written as 
\begin{equation}
{{\displaystyle d^2 \sigma} \over {\displaystyle d E_{K}d \Omega_K}}
=\Bigl( {d \sigma \over d \Omega} \Bigr)^{\rm opt}_{\omega,\theta_K}
S(\omega,\theta_K),
\label{eqn:e6}
\end{equation}
where 
$({d \sigma/d \Omega})^{\rm opt}$ is 
an ``optimal'' $\pi^+$+$n$$\to$$K^+$+$\Lambda$ 
cross section which is defined by
\begin{equation}
\Bigl( {d \sigma \over d \Omega} \Bigr)_{\omega,\theta_K}^{\rm opt}
\equiv { E_K E_\pi \over (2\pi)^2}
{p_K \over p_\pi}|t^{\rm opt}(\omega,\theta_K)|^2, 
\label{eqn:e1}
\end{equation}
and $S(\omega,\theta_K)$ is a strength function for 
hypernuclear production.
Here we assumed to hold a factorization 
in the production cross section 
which is a product of the $\pi^+$+$n$$\to$$K^+$+$\Lambda$
cross section, and of the strength function, 
because we wish to understand the structure of the QF spectrum
separately and distinctively. 
In order to compare directly the inclusive $K^+$ spectrum 
calculated by eq.(\ref{eqn:e6}) 
with the ($\pi^+$,$K^+$) experimental data, 
we need to evaluate the strength function including $\Lambda$ bound, 
resonance and continuum states.
Then we use Green's function method \cite{Morimatsu94}, which is one 
of the most powerful treatments for calculating the strength function.
The $\Lambda$ single-particle potential is well-known phenomenologically
and empirically \cite{Millener88}.
Assuming a Woods-Saxon form,  the $\Lambda$-nucleus potential is
written as $U_\Lambda(r)= V^0_\Lambda [1 + \exp{((r-R)/a)}]^{-1}$, 
where $V_\Lambda^{0}$=$-$28.0 MeV, $a$=0.60 fm, 
$r_0$= 1.128+0.439$A^{-{2/3}}$ fm 
and $R$=$r_0(A-1)^{1/3}$ fm \cite{Millener88}.
For $A$=12 we obtain the $\Lambda$-nucleus potential 
with $r_0$= 1.212 fm and $R$= 2.695 fm. 
We calculate single-particle wave functions for a neutron in the target nucleus, 
using a Woods-Saxon potential \cite{Bohr69} 
and adjusting its strength of $V_N^0$=$-$64.8 MeV to reproduce the data of the 
charge radius of 2.46 fm \cite{Vries86}. 
We input the single-particle energies and widths which are referred 
from ($e$,~$e'p$) reactions for light nuclei \cite{Jacob66},
because deep-hole states for a neutron play significant 
roles in contributing the $\Lambda$ spectrum 
in the QF region \cite{Tadokoro95}.

The full distorted-waves of the $\pi^+$-nucleus and $K^+$-nucleus 
are important to reproduce the absolute value of the cross sections.
Due to a large momentum transfer by the ($\pi^+$,$K^+$) reaction, 
we must calculate partial-waves for high angular-momentum states.
Thus we simplify the computational procedure by using 
the eikonal approximation for distorted waves of meson-nucleus 
states \cite{Dover80,Bouyssy77,Hausmann89}.
We use a matter-density distribution 
fitting to the data for a charge-density distribution \cite{Vries86}, 
and take higher-$\ell$ angular-momentum states 
for $\ell \leq$ 30 sufficiently.

\Figuretable{FIG. 3}

\begin{figure}[thb]
  \begin{center}
  \includegraphics[width=.55\textwidth]{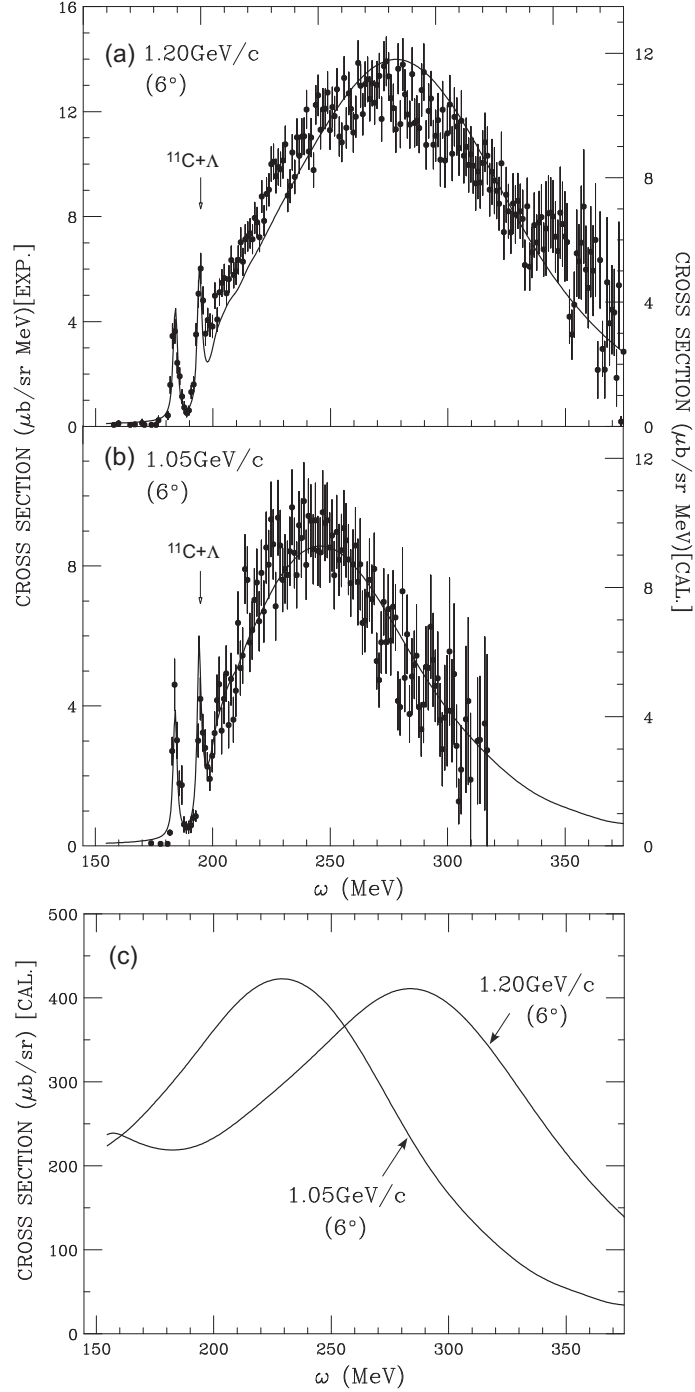}
  \end{center}
  \caption{\label{fig:3}
  Numerical results for $^{12}$C($\pi^+$,$K^+$) reactions, as a function 
  of the energy transfer $\omega$. 
  Calculated inclusive $K^+$ spectra are shown at 
  (a) $p_\pi$=1.20 GeV/c ($\theta_K=$ 6$^\circ$) and (b) 1.05 GeV/c (6$^\circ$), 
  together with the data \cite{Saha03}. 
  The spectra are folded with a detector resolution of 2 MeV FWHM.
  (c) The ``optimal'' cross sections $(d \sigma/d \Omega)^{\rm opt}$
  for the $\pi^+$+$n$$\to$$K^+$+$\Lambda$ reaction on the $^{12}$C target
   are drawn at 1.20 GeV/c (6$^\circ$) and 1.05 GeV/c (6$^\circ$).
  }
\end{figure}

Let us examine the hypernuclear production cross section 
for the ($\pi^+$,$K^+$) reaction on a $^{12}$C target. 
In Fig.~\ref{fig:3}c, we show the optimal $\pi^+$+$n$$\to$$K^+$+$\Lambda$ 
cross sections $(d \sigma/d \Omega)^{\rm opt}$ 
at $p_\pi =$ 1.20 GeV/c ($\theta_K$= 6$^\circ$) and 1.05 GeV/c (6$^\circ$), 
as a function of $\omega$. 
Here we used the elementary $\pi^+$+$n$$\to$$K^+$+$\Lambda$ amplitude 
analyzed by Sotona and $\check{\rm Z}$ofka \cite{Sotona89}.
We find that the $\omega$-dependence of $(d \sigma/d \Omega)^{\rm opt}$ 
is characterized by the incident $\pi^+$ momentum; 
the peak which originates from $N^*$ resonances is located at 
$\omega \simeq$ 280 MeV for $p_\pi=$ 1.20 GeV/c, 
and at $\omega \simeq$ 230 MeV for $p_\pi=$ 1.05 GeV/c. 
The peak position is shifted downward as $p_\pi$ decreases.
Note that the on-energy-shell process 
leads to an appearance of the energy-dependence of 
the $\pi^+$+$n$$\to$$K^+$+$\Lambda$ t-matrix 
in the nucleus \cite{Gurvitz86,Noumi03}.

In Fig.~\ref{fig:3}a, we show the calculated 
inclusive $K^+$ spectrum 
for the $^{12}$C($\pi^+$,$K^+$) reaction at $p_\pi$= 1.20 GeV/c 
($\theta_K$= 6$^\circ$), together with the data \cite{Saha03}.
We find that the shape of the spectrum 
reproduces that of the data overall.
The QF peak is located at $\omega \simeq $ 275 MeV 
which corresponds to about 80 MeV 
above the $^{11}$C$_{\rm g.s.}$+$\Lambda$ threshold, and
its absolute value is compared with the data.
This agreement comes directly from the $\omega$-dependence of 
$(d \sigma/d \Omega)^{\rm opt}$, as seen in Fig.~\ref{fig:3}c.
We also confirm that the contribution of a neutron 
(0s$_{1/2}$)$^{-1}$ state is important in the QF 
spectrum \cite{Tadokoro95}.

The more direct test of the validity of $t^{\rm opt}(\omega,\theta_K)$
is to observe the spectrum under a different incident 
$\pi^+$ momentum \cite{Saha03}. 
As shown in Fig.~\ref{fig:3}b, 
the data of $p_\pi$= 1.05 GeV/c has a peak arising 
at $\omega \simeq$ 240 MeV, and 
its width of about 80 MeV which is extremely narrower than 
the data of $p_\pi=$ 1.20 GeV/c.
This sizeable change is finely overcome in our calculations;
the resultant spectrum can reproduce fully the data 
including from $\Lambda$ bound to QF regions. 
A peak position in the QF spectrum is also in excellent 
agreement with that of the data, and its absolute value 
is quite good. 
Due to a change to $p_\pi =$ 1.05 MeV/c, 
the peak position of $(d \sigma/d \Omega)^{\rm opt}$ 
is shifted downward by about 45 MeV, as shown in Fig.~\ref{fig:3}c.
Then its behavior makes the width of the QF spectrum look narrow, 
as shown in Fig.~\ref{fig:3}b.
Consequently, we show that the calculated spectra in eq.(\ref{eqn:e6})
can explain all the data in $\Lambda$ bound and QF regions simultaneously, 
and clarify that $t^{\rm opt}(\omega,\theta_K)$ enables 
us to describe the fine spectrum for the ($\pi^+$,$K^+$) reaction.
The results suggest that if we evaluate the data of the QF spectra
to understand properties of hyperon- or meson-nucleus interactions, 
we need careful consideration for the energy-dependence of 
the elementary cross section. 

In a more quantitative comparison, 
productions for the $\Lambda$ ground state agree with the data at 
1.20 GeV/c and 1.05 GeV/c, 
whereas production for the QF spectrum at 1.20 GeV/c seems to be 
slightly smaller than the data by about 15\%.
This might suggest a possibility of $\Lambda$ QF 
production via $\Sigma$ components
because a $\Sigma$ channel is already opened at 1.20 GeV/c. 

In conclusion, we have proposed
the optimal Fermi-averaging for the elementary 
$\pi^+$+$n$$\to$$K^+$+$\Lambda$ t-matrix in the ($\pi^+$,$K^+$) reactions 
on the nuclear target. 
The resultant spectra for the $^{12}$C($\pi^+$,$K^+$) reactions 
can explain the experimental data very well,
because the $\omega$ energy-dependence originates from the nature of 
the optimal Fermi-averaging t-matrix.
The on-energy-shell $\pi^+$+$n$$\to$$K^+$+$\Lambda$ processes
in the nucleus lead to a success in describing the 
($\pi^+$,$K^+$) spectrum.
Our treatment would give us a powerful way to calculate 
a spectrum beyond the ordinary DWIA \cite{Harada04}.

The authors are obliged to  Professor H. Noumi, Dr. P.K. Saha 
and Professor M. Kawai for variable discussion. 
One of the authors (T.H) thanks to Professor Y. Akaishi, Professor A. Gal
and  Professor H. Tanaka for useful comments. 
This work was supported by JSPS, 
the Grant-in-Aid for Scientific Research (C), No. 13640302.


\end{document}